\documentclass[a4paper, 10pt]{article} 
\usepackage[utf8]{inputenc}
\usepackage{authblk}
\usepackage{amssymb}
\usepackage{amsmath}
\usepackage{amsfonts}
\usepackage{mathrsfs}
\usepackage{pxfonts}
\usepackage{latexsym}
\usepackage{pgfplots} 
\usepackage{graphicx}

\usepackage[breaklinks=true,colorlinks=true,linkcolor=blue,urlcolor=blue,citecolor=blue]{hyperref}

\begin{document}

\title{Canonical gravity and scalar fields}
\author{Avadhut Purohit}

\maketitle

\begin{abstract}
A combined variable theory of gravity and scalar field is developed under ADM formulation by redefining new canonical 
conjugate variables $(\Phi,\Pi)$. It is a dynamical theory of the space-matter with implicit time dependence. Quantizing
this new field theory results in quantization of geometry and scalar field combined. 
\end{abstract}

\newpage
\section{Introduction}
 \begin{center}
  \textbullet \textbf{Motivation}: 
 \end{center}
 \hspace*{1 cm} General Relativity and Quantum Field Theory are two most successful theories of $19^{\text{th}}$ century. 
General Relativity is a classical theory of dynamical space-time. Quantum theory tells us that every dynamical 
field has quantum properties. But General Relativity being a non-renormalizable theory is very difficult to quantize.
 Whereas, Quantum Field Theory is built on a fixed background and relies on Poincar\'{e} invariance. It is a quantum 
theory of matter fields. But General Relativity tells us that space-time is also dynamical. Therefore, both theories are 
still incomplete. \\
\hspace*{1 cm} Anorwitt, Deser and Misner [1] formulated a theory which brings gravity closer to Lorentz covariant field 
theories by disentangling dynamical degrees of freedom from the gauge variables. Loop Quantum Gravity is one of the most 
promising amongst canonical theories of gravity. It is a dynamical theory of connections ([2] and [3] can be referred
for further details). A granularity of space is a result of quantization program. But the theory is far from being a 
 complete theory for various reasons such as the existence of semi-classical limit is not proven so far ([4] section VI, A
 summarizes some more open problems as well as achievements of the theory). There is an important take away from the 
 theory that is \textit{fields are inter-dependent and evolve with respect to one another}.     
\begin{center}
 \textbullet \textbf{Approach of this paper}:
\end{center}
\hspace*{1 cm} The history of Lorentz covariant theories has already taught us a lesson that \textit{we need a concept of a
`field' in order to unify Special Relativity with Quantum Mechanics}. General Relativity is a generalization of Special 
Relativity. Hence, quantization of General Relativity requires a concept of a `field' as well. 
It is already noted above that `fields evolve with respect to one another'. Time is not a direct observable (in a sense 
that we measure time in terms of relative displacements between matter fields) in contrast to space and matter. 
Thus, the theory developed here is a dynamical theory of the space-matter with implicit time dependence.\\ 
\hspace*{1 cm} ADM form of gravity is taken as a starting 
point which not only brings gravity closer to Lorentzian field theories but also singles out time thereby allowing to build a 
dynamical theory of gravity-scalar field. After separating spatial and temporal parts of fields, the full Hamiltonian 
$H_{\text{total}}=0$ is reinterpreted as an equation of motion for the new combined variable field. This new field theory is 
then quantized canonically. Quantization of gravity and scalar field combined is a result of the quantization procedure. \newline 
\newline
Note: Throughout the paper I am going to work in the units of $16\pi G =1$, $\hbar =1$ and $c=1$. Wherever required these can be 
plugged in by dimensional analysis. 

 \section{Hamiltonian form of gravity and scalar field theory}
\hspace*{1 cm} Action for gravity together with the scalar field is taken and carried out 3+1 decomposition by foliating 
4-dimensional space-time manifold into one parameter family of hypersurfaces $\Sigma_{t}$. Any time-like vector field $T^{\mu}$
decomposed into $n^{\mu}$ unit vector normal to hypersurface and $N^{\mu}$ shift vector tangential to hypersurface. $N$ is 
lapse function. (Note: $\mu,\nu..$ represents space-time indices and $a,b,c..$ represents spatial indices.) 
\begin{equation}
 T^{\mu}(t, \vec{x}) \coloneqq N(t, \vec{x}) n^{\mu}(t, \vec{x}) + N^{\mu}(t, \vec{x}) 
\end{equation}
Detail discussion of 3+1 decomposition or ADM formulation can be found in references [5], section I.2.1 as well as in the 
book [6], section I, chapter 1. 
\begin{equation}
 \mathscr{A} =  \int_{\mathbb{R}}dt \int_{M} d^3 x \left( |N| \sqrt{q} \left\lbrace  \left( R^{(3)}+
  K^{ab}K_{ab} - K^2 \right) + 
   \left( \frac{1}{2} \dot{\phi}^{2} - \frac{1}{2} |\nabla \phi|^{2} - V(\phi) \right) \right\rbrace  \right) (t,\vec{x}) 
 \end{equation}
 Curvature in 4-dimensional space $R^{(4)}$ decomposed into $R^{(3)}+ K^{ab}K_{ab} - K^2$. $K_{ab}(t,\vec{x})$ is extrinsic 
 curvature of hypersurface $\Sigma_{t}$, $R^{(3)}$ is a spatial curvature and $q_{ab}(t,\vec{x})$ is 3-metric both 
 defined on 3-manifold and $\sqrt{q} = \sqrt{\text{det}(q_{ab})}(t,\vec{x})$. Canonical conjugate momenta corresponding 
 to $q_{ab}(t,\vec{x})$, $N^{a}(t,\vec{x})$, $N(t,\vec{x})$ and $\phi(t,\vec{x})$ are respectively (derived in [6], section 
 1.2, equation 1.2.1),
 \begin{align}
 & P^{ab}(t,\vec{x}) \coloneqq \left( \frac{|N|}{N} \sqrt{q} \left( K^{ab} - K q^{ab} \right) \right) 
 (t,\vec{x}) \\
  & C_{a}(t,\vec{x}) \coloneqq \Pi_{a}(t,\vec{x}) = 0 \\
  & C(t,\vec{x}) \coloneqq \Pi(t,\vec{x}) =0 \\
  & P_{\phi}(t,\vec{x}) \coloneqq \left( |N| \sqrt{q} \dot{\phi} \right)(t,\vec{x}) 
 \end{align}
 $\left( q_{ab},P^{cd},\phi,P_{(\phi)},N,\Pi,N^{a},\Pi_{a}\right) (t,\vec{x})$ 
 forms a phase space and corresponding only non-trivial Poisson Brackets (borrowed from [5], 
 equation 1.2.1.9) are given as, 
\begin{align}
 & \left\lbrace q_{ab}(t,x),  P^{cd}(t,x^{\prime}) \right\rbrace = 
 \delta^{c}_{(a} \delta^{d}_{b)}  \delta^{(3)}(x,x^{\prime}) \\
 &  \left\lbrace  \phi(t,x), P_{(\phi)}(t,x^{\prime}) \right\rbrace = \delta^{(3)}(x,x^{\prime})
\end{align}
Lagrangian for gravity (refer I.2.1.7 of [5] or 1.2.5 of [6]) and for matter field is given as,
\begin{align}
 & L_{\text{grav}} =  \int_{M} d^3x \left\lbrace P^{ab}\dot{q}_{ab}+\Pi^{a}\dot{N}_{a}+\Pi\dot{N} - 
\left( \lambda C+\lambda^{a}C_{a}+N^{a}H_{a}+|N| H_{\text{scalar}} \right) \right\rbrace (t,\vec{x}) \\
 & L_{\phi} = \int_{M} d^3x \left\lbrace P_{\phi}\dot{\phi}- |N| \sqrt{q} \left( \frac{P^{2}_{\phi}}{2N^2 
\text{det}(q)} + \frac{1}{2}|\nabla\phi|^{2} + V(\phi) \right) \right\rbrace (t,\vec{x})
\end{align}
 $\lambda$ and $\lambda^{a}$ are Lagrange undetermined multipliers and there are no equations to solve for them. 
 Therefore, reduced constrained action (refer I.2.1.13 of [5] or 1.3.1 of [6]) is written as 
\begin{equation}
 \mathscr{A} = \int_{\mathbb{R}}dt  \int_{M} d^3x \left\lbrace P^{ab}\dot{q}_{ab} + P_{\phi}\dot{\phi} - 
 \left( N^{a}\mathscr{H}_{a}+|N| \mathscr{H}_{\text{scalar}} + |N|\mathscr{H}_{\phi} \right)\right\rbrace (t,\vec{x})
\end{equation}
\begin{equation}
  \mathscr{H}_{a} \coloneqq \left( -2 q_{ac}D_{b}P^{bc} \right) (t,\vec{x})
\end{equation}
$D_{b}$ is unique torsion-free covariant differential compatible with $q_{ab}$. 
\begin{align}
 & \mathscr{H}_{\text{scalar}} \coloneqq \left( \frac{1}{\sqrt{q}} \left( q_{ac}q_{bd}-
 \frac{1}{2}q_{ab}q_{cd} \right) P^{ab}P^{cd} - \sqrt{q} R^{(3)} \right) (t,\vec{x}) \\
 &\mathscr{H}_{\phi} = \sqrt{q} \left( \frac{P^{2}_{\phi}}{2N^2 
\text{det}(q)} + \frac{1}{2}|\nabla\phi|^{2} + V(\phi) \right) (t,\vec{x})
\end{align}
$ \mathscr{H}_{a}$ are called `Diffeomorphism constraints' or vector constraints. These constraints generate diffeomorphism 
on 3-Manifold. $\mathscr{H}_{\text{scalar}}$ `Hamiltonian constraint' or scalar constraint generates temporal gauge 
transformation (refer 1.2.6 of [6] or I.2.1.10 of [5]). Symmetrise $\mathscr{H}_{\text{scalar}}$ in 
$a\leftrightarrow b$ and $c\leftrightarrow d$ 
\begin{equation}
\mathscr{H}_{\text{scalar}} = \left( -\frac{1}{2} f_{abcd} P^{ab}P^{cd} - \sqrt{q} R^{(3)}\right) (t,\vec{x})
\end{equation}
Where $f_{abcd}$ is defined as,
\begin{equation}
 f_{abcd} \coloneqq \left( \frac{1}{\sqrt{q}} \left(-q_{bc}q_{ad}-q_{ac}q_{bd}+q_{ab}q_{cd} \right) \right) (t,\vec{x})
\end{equation}
\begin{equation} \label{eqn37}
H_{\text{scalar}} = \int_{M} d^3x  |N| \left(  - \frac{1}{2} f_{abcd} P^{ab}P^{cd}- \sqrt{q} R^{(3)} \right)
(t,\vec{x}) 
\end{equation}
\begin{equation}
 H_{\text{vector}} = \int_{M} d^3x N^{a}\left( -2 q_{ac}D_{b}P^{bc} \right) (t,\vec{x}) 
\end{equation}
Both these constraints are first class constraints. These constraints are preserved under evolution. (Section I.2.1 of [5] 
can be referred for further detailed analysis)
Separate time and space dependent parts of $q_{ab}(t,\vec{x})$ and $P^{ab}(t,\vec{x})$ as 
\begin{align}
 & q_{ab}(t,\vec{x}) \rightarrow q_{a}(t) \mathring{q}_{ab}(\vec{x}) \hspace{0.5 cm}
 & P^{ab}(t,\vec{x}) \rightarrow P^{a}(t) \mathring{P}^{ab}(\vec{x})
\end{align}
First term in the scalar Hamiltonain constraint becomes,
\begin{align}
 & \int_{M} d^3x |N(t)|f_{abcd}(t,\vec{x})P^{ab}(t,\vec{x})P^{cd}(t,\vec{x}) \nonumber \\
 & =|N(t)| \eta_{ac}(t)P^{a}(t)P^{c}(t) 
 \int_{M} d^3x \left( \mathring{f}_{abcd}(\vec{x})\mathring{P}^{ab}(\vec{x})\mathring{P}^{cd}(\vec{x}) \right) \nonumber\\
 & =\eta_{ac}(t)P^{a}(t)P^{c}(t) \label{eqn5} 
\end{align}
With, 
\begin{align}
 & \mathring{f}_{abcd}(\vec{x}) = \frac{1}{\sqrt{\mathring{q}}(\vec{x})} \left( \mathring{q}_{ab}(\vec{x})
 \mathring{q}_{cd}(\vec{x})-
 \mathring{q}_{ac}(\vec{x})\mathring{q}_{bd}(\vec{x})-\mathring{q}_{bc}(\vec{x})\mathring{q}_{ad}(\vec{x}) \right)\nonumber \\
 & \eta_{ac}(t)\coloneqq q_{a}(t)q_{c}(t) \label{eqn22} \\
 & v = \int_{M} d^3x  \left( \mathring{f}_{abcd}(\vec{x}) \mathring{P}^{ab}(\vec{x}) \mathring{P}^{cd}(\vec{x}) \right) \label{eqn6} \\
 & |N(t)|= \frac{\sqrt{q}(t)}{v}  \label{eqn7} 
\end{align}
Reason to choose such $|N(t)|$ is to avoid the term $\sqrt{q}(t)$ appearing in the denominator of $f_{abcd}$. This choice will 
give the kind of full Hamiltonian constraint which will allow us to construct a suitable Lagrangian for the new field which
is done in the next section. Diffeomorphism constraints are given as 
\begin{equation} 
 -2 \int_{M} d^3x N^{a}\left( q_{ac}D_{b}P^{bc} \right)(t,\vec{x}) = -2 q_{i}(t)P^{i}(t) \hspace{0.2 cm} N^{a}
 \int_{M} d^3x  \left( \mathring{q}_{ac}(\vec{x})D_{b}(\vec{x})\mathring{P}^{bc}(\vec{x}) \right)  
 \end{equation}
Combined variable field theory (section 3) requires $N^{a}$ to be chosen in such a way that 
\begin{equation} \label{eqn8} 
 N^{a}\int_{M} d^3x  
\left( \mathring{q}_{ac}(\vec{x})D_{b}\mathring{P}^{bc}(\vec{x}) \right)=-i
\end{equation}
This choice of shift vector will enable us to find out Lagrangian of combined variable theory. 
\begin{equation} \label{eqn9} 
 H_{\text{vector}}= 2i q_{k}P^{k}
\end{equation}
The second term in the $H_{\text{scalar}}$ becomes
\begin{equation} \label{eqn10} 
   V(\vec{q}(t)) =\text{det}(q_{k}(t)) \frac{\int_{M} d^{3}x \sqrt{\mathring{q}} R^{(3)}(t,\vec{x})}{\int_{M} d^3x 
  \left( \mathring{f}_{abcd}(\vec{x}) \mathring{P}^{ab}(\vec{x}) \mathring{P}^{cd}(\vec{x}) \right)} 
\end{equation}
By collecting (\ref{eqn5}), (\ref{eqn9}) and (\ref{eqn10}) we get gravitational
part of the full Hamiltonian (`t' dependance is supressed)
\begin{equation} \label{eqn16}
 H_{\text{scalar}}+H_{\text{vector}} = - \frac{1}{2}\eta_{ij} P^{i}P^{j}+2i q_{k}P^{k}-V(\vec{q})
\end{equation}
$\eta_{ij}$ is real and symmetric matrix
\begin{align} \label{eqn34} 
 \eta_{ij} = 
 \begin{pmatrix}
  q_{1}^{2} &q_{1}q_{2} &q_{1}q_{3} \\
  q_{2}q_{1} &q_{2}^{2} &q_{2}q_{3} \\
  q_{3}q_{1} &q_{3}q_{2} &q_{3}^{2}
 \end{pmatrix}
\end{align}
 Let us now carry out similar analysis to the matter (scalar field) part of the full 
Hamiltonian 
\begin{equation} \label{eqn11}
 H_{\text{matter}} = \int_{M} d^3x |N| \left( \frac{P^{2}_{\phi}}{2N^{2}\sqrt{q}} 
 + \frac{\sqrt{q}}{2}|\nabla\phi|^{2} +\sqrt{q} V(\phi) \right) (t,\vec{x})
\end{equation}
separate the temporal and spatial parts of the matter field and its canonical conjugate momentum $\phi(t,\vec{x})=\phi(t)
\mathring{\phi}(\vec{x})$, $P_{\phi}(t,\vec{x})=P_{\phi}(t)\mathring{P}_{\phi}(\vec{x})$ and then carry out integration 
over spatial part we get,
\begin{equation}
 \frac{1}{2} \int_{M} d^3x \frac{|N|}{N^{2}\sqrt{q(t)}\sqrt{\mathring{q}(\vec{x})}} 
 P^{2}_{\phi}(t)\mathring{P}^{2}_{\phi}(\vec{x}) = \frac{vv^{\prime}}{2q(t)}P^{2}_{\phi}(t)
\end{equation}
 With $v= \int_{M} d^3x  \left( \mathring{f}_{abcd}(\vec{x}) \mathring{P}^{ab}(\vec{x})
 \mathring{P}^{cd}(\vec{x}) \right)$ and $v^{\prime} = \int_{M} d^3x \frac{ \mathring{P}^{2}
 (\vec{x})}{\sqrt{\mathring{q}}}$. Redefining $P_{\phi}(t)$ and $\phi(t)$ such that Poisson Bracket remain unchanged.
\begin{align} \label{eqn12}
&   \sqrt{\frac{q(t)}{vv^{\prime}}} P_{\phi}(t) \rightarrow P_{\phi} (t) \hspace{0.5 cm}
 &  \sqrt{\frac{vv^{\prime}}{q(t)}} \phi(t) \rightarrow  \phi(t) 
\end{align}
\begin{equation} \label{eqn13}
 \left\lbrace \phi(t), P_{\phi}(t) \right\rbrace = 1 
\end{equation}
This redefinition makes
\begin{equation} \label{eqn14}
 \text{first term of }H_{\text{matter}} = \frac{1}{2}P^{2}_{\phi}(t)
\end{equation}
due to the redefinition $|\nabla \phi|^{2}$ and massive coupling term 
\begin{equation}
   \frac{1}{2} \int_{M} |N| \sqrt{q_{ab}} |\vec{\nabla} \phi|^{2}(t,\vec{x}) \rightarrow  \frac{1}{2}\phi^{2}(t) 
   \left( \int_{M}  d^{3}x v^{\prime} \sqrt{\mathring{q}_{ab}(\vec{x})} \phi(\vec{x}) \nabla^{2} \phi(\vec{x}) \right)
\end{equation}
\begin{equation}
 \frac{1}{2} m^{2} \int_{M} d^{3}x |N| \sqrt{q_{ab}} \phi^{2}(t,\vec{x}) \rightarrow 
 \frac{1}{2} m^{2} \phi^{2}(t)
  \left( \int_{M}d^{3}x \frac{v v^{\prime}}{\text{det }q} \sqrt{\mathring{q}_{ab}(\vec{x})} \phi^{2}(\vec{x}) \right) 
\end{equation}
These two equations can be added up and spatial dependance is absorbed into mass. This coupling is a functional of $\vec{x}$.
\begin{equation}
  \int_{M}  d^{3}x v^{\prime} \sqrt{\mathring{q}_{ab}(\vec{x})} \phi(\vec{x}) \nabla^{2} \phi(\vec{x}) + m^{2}
 \int_{M}d^{3}x \frac{v v^{\prime}}{\text{det }q} \sqrt{\mathring{q}_{ab}(\vec{x})} \phi^{2}(\vec{x}) \rightarrow \mu^{2}
\end{equation}
Similar analysis can also be carried out and spatial parts can be absorbed into respective couplings. Then matter Hamiltonian
becomes
\begin{equation} \label{eqn17}
 H_{\text{matter}}= \frac{1}{2}P^{2}_{\phi} + V(\phi,q)
\end{equation}
The full Hamiltonian is a sum of (\ref{eqn16}) and (\ref{eqn17}),
\begin{equation} \label{eqn18}
 H_{\text{total}}=\frac{1}{2}P^{2}_{\phi} - \frac{1}{2}\eta_{ij} P^{i}P^{j}+2i q_{k}P^{k}
 -V(\vec{q})+V(\phi,q)
\end{equation}
$(q_{i},P^{j},\phi,P_{\phi})$ is reduced phase 
space for $ H_{\text{total}}$ with only non-trivial Poisson brackets $\left\lbrace q_{i},P^{j}\right\rbrace=\delta^{j}_{i}$ 
 and $\left\lbrace \phi,P_{\phi}\right\rbrace=1$. Equations of motion are given by 
\begin{align}
 & \dot{P_{\phi}}(t) = \left\lbrace P_{\phi}(t) , H_{\text{total}} \right\rbrace
 & \dot{\vec{P}}(t) = \left\lbrace \vec{P}(t) , H_{\text{total}} \right\rbrace \\
 & \dot{\vec{q}}(t) = \left\lbrace \vec{q}(t) , H_{\text{total}} \right\rbrace 
 & \dot{\phi}(t) = \left\lbrace \phi(t) , H_{\text{total}} \right\rbrace
\end{align}
We have 3 configuration variables for gravity with one first class constraint. First class constraint reduces configuration
degree of freedom by 1. These are degrees of freedom of gravity. \newline \newline 
\textit{case 1:} In special relativistic limit, $R^{(3)}=0$ imply $V(\vec{q})=\text{const.}$ and extrinsic curvature 
($K_{ab}=0$) imply $P^{j}=0$. Then, $$H_{\text{total}}=\frac{1}{2}P^{2}_{\phi} + V(\phi)$$ Note that $V(\phi,q)=V(\phi)$ 
because $q(t)=1$. This is exactly the Hamiltonian for scalar field with a spatial part being integrated over and contribution
of that part is absorbed inside coupling constants. This Hamiltonian gives time evolution. \newline \newline
\textit{case 2:} In zero scalar field limit, $V(\phi,q)=0$. Equation of motion give $\dot{\phi}=P_{\phi}$ but since $\phi=0$ 
and $\dot{\phi}=0 \rightarrow P_{\phi}=0$. Therefore, $$H_{\text{total}}=- \frac{1}{2}\eta_{ij} P^{i}P^{j}+2i 
q_{k}P^{k} -V(\vec{q})$$ This is the Hamiltonian for gravity with a lapse function and a shift 
vector chosen appropriately. Hamiltonian does not give physical time evolution. Instead, it gives a gauge transformation. \\
\section{Combined variable theory of gravity and scalar field}
\subsection{Classical Theory}
Rewrite (\ref{eqn18}) by using $P_{\phi}=-i\frac{\partial}{\partial\phi}=-i\partial_{\phi}$, 
$P^{k}=-i\frac{\partial}{\partial q_{k}}=-i\partial^{k}$. Combined potential is defined as
$\frac{1}{2}V(q_{k},\phi) \coloneqq V(\vec{q})-V(\phi,q)$ is combined potential energy of gravity and scalar field.
\begin{align*}
 \hat{H}_{\text{total}}\Phi = \left( -\frac{1}{2}\frac{\partial^{2}}{\partial\phi^{2}} + \frac{1}{2}\eta_{ij}
 \frac{\partial}{\partial q_{i}}\frac{\partial}{\partial q_{j}} + 
 2  q_{i}\frac{\partial}{\partial q_{j}}- \frac{1}{2}V(q_{k},\phi) \right) \Phi
\end{align*}
[Note:`t' dependence is suppressed from here onwords.] This equation is now interpreted as an equation of motion for classical
field $\Phi$ of combined variables.
\begin{equation}
  \left( \frac{\partial^{2}}{\partial\phi^{2}}-\frac{\partial}{\partial q_{i}}
 \eta_{ij} \frac{\partial}{\partial q_{j}}+V(q_{k},\phi) \right) \Phi =0
\end{equation}   
 $\frac{\partial}{\partial q_{i}}\eta_{ij}\frac{\partial}{\partial q_{j}} = \eta_{ij}\frac{\partial}{\partial q_{i}} 
\frac{\partial}{\partial q_{j}}+ \frac{\partial \eta_{ij}}{\partial q_{i}}\frac{\partial}{\partial q_{j}}$. In the second term,
$\frac{\partial}{\partial q_{i}}$ acting on $q_{i}$ of $\eta_{ij}$ gives $3q_{j}$ and acting on $q_{j}$ of $\eta_{ij}$ gives
1 $q_{j}$. This second term is a result of Diffeomorphism constraints. If we had chosen a different shift vector, that would 
have introduced extra derivative coupling but the choice (\ref{eqn8}) compensates this extra derivative term. 
Taking plane wave solution $\Phi\sim e^{i(P_{\phi}\phi+P^{k}q_{k})}$ we recover (\ref{eqn18}) which has special 
relativistic as well as zero scalar field limit. \\
Explicit form of combined potential depends on both gravity and scalar field coupllings  
\begin{align} \label{eqn40}
 &V(q_{k},\phi) =  2 \left( \left(\text{det }q \right) \frac{ \int_{M} d^{3}x \sqrt{\mathring{q}} R^{(3)}}{\int_{M} d^3x  
 \left( \mathring{f}_{abcd}(\vec{x}) \mathring{P}^{ab}(\vec{x}) \mathring{P}^{cd}(\vec{x}) \right)} \right)\\ \nonumber 
 - & 2 \left( \epsilon_{0} \sqrt{\text{det } q} \phi
 + \mu^{2}\phi^{2}+ \frac{\beta}{\sqrt{\text{det } q}}\phi^{3} +\frac{\lambda}{ (\text{det }q)}\phi^{4} \right)
\end{align}
All couplings are functional of $\vec{x}$. $\eta_{ij}$ serves as a spatial metric for the space of $(\phi,\vec{q})$. 
Action for this combined variable field is given as 
\begin{equation} \label{eqn20}
 \mathscr{A} =  \int d\phi \int d^{D} q \left\lbrace \frac{1}{2}(\partial_{\phi} \Phi)^{2}-
 \frac{1}{2}(\partial^{i}\Phi)\eta_{ij}(\partial^{j}\Phi) -\frac{1}{2} V(q_{k},\phi)\Phi^{2} \right\rbrace 
\end{equation}
$D=0,1,2 \text{ or } 3$ depending on how many components of $\vec{q}$ are time dependent. Both the gravity as well as the scalar 
field contribute to the massive coupling $V(q_{k},\phi)$ of the combined variable field. It can also be noted that gravity couples
differently with different kinds of scalar field couplings.  
For $V<0$ and $\alpha>0$ 
\begin{align} \label{eqn35}
 \mathscr{A} =  \int d\phi \int d^{D} q \left\lbrace \frac{1}{2}(\partial_{\phi} \Phi)^{2}-
 \frac{1}{2}(\partial^{i}\Phi)\eta_{ij}(\partial^{j}\Phi) -\frac{1}{2} V \Phi^{2}-\frac{1}{4}\alpha \Phi^{4} \right\rbrace 
\end{align}
The field is unstable at $\Phi=0$ and will condense into a stable state 
$\Phi \mid_{\text{vac}}=\pm \sqrt{\frac{-V}{\alpha}}$.
 Symmetry of the ground state is spontaneously broken. If we shift 
$\Phi = \rho - \Phi \mid_{\text{vac}}$, we get
\begin{align} \label{eqn36}
 -\frac{1}{2} V \Phi^{2}-\frac{1}{4}\alpha \Phi^{4} \longrightarrow \frac{1}{2}2V\rho^{2}-\frac{1}{4}\alpha \rho^{4}
 + \left( V\Phi \mid_{\text{vac}}+\alpha \Phi \mid_{\text{vac}}^{3}\right)\rho +\alpha \Phi \mid_{\text{vac}}\rho^{3}
\end{align}
Extra constant terms are not written as they are inconsequential. $\rho^{2}$ term comes with a correct sign and therefore 
mass spectrum remain positive definite. Treatment of this theory is beyond scope of this paper and will be discussed in the 
next paper. \newline
\hspace*{1 cm}A 4-dimensional space-matter manifold is equipped with a metric
\begin{equation} \label{eqn21}
 g_{\mu \nu} \coloneqq 
 \begin{pmatrix}
   1  & 0 \\ 0 & -\eta_{ij} 
 \end{pmatrix}
\end{equation}
which allows us to measure the length in the space of $(\phi,\vec{q})$
\begin{equation} \label{eqn26}
 dl^{2} = g_{\mu\nu} dq^{\mu}dq^{\nu} = d\phi^{2} - \eta_{ij}dq^{i}dq^{j}
\end{equation}
Rewrite action (\ref{eqn20}) in covariant form
\begin{equation} \label{eqn23}
 \mathscr{A} = \int d\phi \int d^{D} q \left\lbrace \frac{1}{2} g_{\mu\nu}\partial^{\mu}\Phi 
 \partial^{\nu}\Phi -\frac{1}{2} V(q_{k},\phi)\Phi^{2} \right\rbrace 
\end{equation}
 $\mu,\nu = 0,1,2,3$ with $0^{\text{th}}$ term being $\phi$ index and $1,2,3$ are metric indices. \newline
 $\partial^{\mu}= \frac{\partial}{\partial q_{\mu}}= \left( \frac{\partial}{\partial \phi}, \frac{\partial}{\partial q_{i}}
 \right)$.
Define \begin{equation} \label{eqn24}
 \Pi_{\mu} \coloneqq \frac{\partial \mathscr{L}}{\partial (\partial^{\mu}\Phi)}= g_{\mu\nu}\partial^{\nu}\Phi
\end{equation} 
Extremizing action under infinitesimal variation of combined variable field $\Phi \rightarrow \Phi + \delta\Phi $ we get 
\begin{equation} \label{eqn25}
  - \int d\phi \int d^{D}q \left( \partial^{\mu}g_{\mu\nu}\partial^{\nu}\Phi + V(q_{k},\phi)\Phi \right)
  \delta\Phi + \int d\phi \int d^{D}q \partial^{\mu}
  \left( g_{\mu\nu}\partial^{\nu}\Phi \hspace*{1 mm}\delta\Phi \right) =0
\end{equation}
Second term is a surface term and if variations are such that $\delta\Phi\rightarrow 0$ on the surface then we get 
(\ref{eqn19}). Equivalently, current $(J_{\mu}\coloneqq g_{\mu\nu}\partial^{\nu}\Phi \hspace*{1 mm}\delta\Phi)$ is conserved 
if equation of motion is satisfied. Invariance of action under $q_{\mu}\rightarrow q_{\mu} + \delta q_{\mu}$ results into 
\begin{align*} \label{eqn26}
\partial^{\rho}T^{\mu}_{\rho} = \partial^{\rho} \left( \frac{\partial \mathscr{L}}{\partial 
(\partial^{\rho}\Phi)}\partial^{\mu}\Phi  - \mathscr{L}\delta^{\mu}_{\rho} \right) =0 
\end{align*}
This gives us four constants of motion. $\mu =0$ gives energy and other three are momentum of the combined variable field
\begin{equation} \label{eqn27}
 E =  \int d^{D}q \left( \frac{\partial \Phi}{\partial \phi}\frac{\partial \Phi}{\partial \phi} + 
 \frac{\partial \Phi}{\partial q_{i}}\eta_{ij}\frac{\partial \Phi}{\partial q_{j}} + V(q_{k},\phi) \Phi^{2} \right)
\end{equation}
\begin{equation} \label{eqn28}
 \mathbf{P}_{i} =  \int d^{D}q \hspace*{2 mm} \frac{\partial \Phi}{\partial 
 \phi}\frac{\partial \Phi}{\partial q_{i}}
\end{equation}
Hamiltonian is obtained by carrying out Legendre transformation 
\begin{equation} \label{eqn29}
 \textbf{H} = \int d^{D}q \left( \frac{1}{2} \Pi^{2} + \frac{1}{2} \frac{\partial \Phi}{\partial q_{i}}
 \eta_{ij} \frac{\partial \Phi}{\partial q_{j}} + \frac{1}{2}V(q_{k},\phi) \Phi^{2} \right)
 \end{equation}
 $\Pi$ ($\Pi_{0}=\frac{\partial \Phi}{\partial \phi}$ of (\ref{eqn24})) is a canonical conjugate momentum corresponds to the 
 field $\Phi$. 
\begin{align} 
 & \left\lbrace \Phi(\phi,\vec{q}), \Pi(\phi,\vec{q}^{\prime})\right\rbrace \coloneqq
 \delta(\vec{q},\vec{q}^{\prime}) \nonumber \\
 & \left\lbrace \Phi(\phi,\vec{q}), \Phi(\phi,\vec{q}^{\prime})\right\rbrace \coloneqq 0 \hspace{0.5 cm}
  \left\lbrace \Pi(\phi,\vec{q}), \Pi(\phi,\vec{q}^{\prime})\right\rbrace \coloneqq 0
\end{align}
 Dynamics of the theory is given by following equations of motion. 
\begin{align} \label{eqn30}
 \frac{\partial \Phi}{\partial \phi} = \left\lbrace \Phi, \textbf{H} \right\rbrace , \hspace{0.5 cm}
  \frac{\partial \Pi}{\partial \phi} = \left\lbrace \Pi, \textbf{H} \right\rbrace
\end{align}
Since, 
\begin{align*}
 \frac{\partial \Phi (\vec{q}^{\prime})}{\partial \phi} =\lbrace \Phi(\vec{q}^{\prime}), \textbf{H} \rbrace 
 &= \int d^{D}q \hspace*{1 mm} \Pi(\vec{q})\delta(\vec{q},\vec{q}^{\prime}) \\
 &=\Pi(\vec{q}^{\prime})
\end{align*}
evolution is consistent. The second equation gives evolution of $\Pi$ with $\phi$. \newline \newline
 \textbullet \textbf{Remarks:} \newline 
\hspace*{1 cm}In order to see physics of classical combined variable field theory, let us first revisit a transition  
 \begin{align*}
     \text{particle} \rightarrow \text{field}
 \end{align*}
I. $H=\frac{p^{2}}{2m}+V(\vec{x},t)$ is the Hamiltonian for a particle or system of particles. Hamiltonian equations give 
trajectory (path) of a particle or system of particles. \newline \newline
II. $\hat{H}\phi=\left( \frac{\hat{p}^{2}}{2m}+\hat{V}(\vec{x},t)\right) \phi$ is reinterpreted as an equation of motion for 
classical field which is defined over space and evolves with time. The second term is seen as coupling which is 
$(\vec{x},t)$ dependent. \newline 
\hspace*{1 cm} Now, examine a transition 
\begin{align*}
 \text{field} \rightarrow \text{combined variable field}
\end{align*}
III. In the case of gravity, space-time itself is a dynamical entity. Hamiltonian equations of motion tell us how $\phi$ varies 
with 3-metric or equivalently how $q_{i}$ changes with $\phi$.  \newline \newline
IV. Unlike case III, combined variable field $\Phi$ is spread over the space of $\vec{q}(t)$. Hamiltonian (\ref{eqn29}) 
gives $\phi$ evolution. Mass of the combined variable field depends not only on the scalar field but also depends on the gravity. 


\subsection{Quantum Theory}
Theory is quantized by raising Poisson brackets of combined variables to commutators 
 \begin{align} \label{eqn1}
  \left[ \hat{\Phi} (\phi, \vec{q}), \hat{\Phi}(\phi, \vec{q}^{\prime}) \right] &= 
   \left[ \hat{\Pi}(\phi, \vec{q}) , \hat{\Pi}(\phi, \vec{q}^{\prime}) \right] = 0  \\
   \left[ \hat{\Phi} (\phi, \vec{q}), \hat{\Pi}(\phi, \vec{q}^{\prime}) \right] &=
   i \hspace*{0.1 cm} \delta(\vec{q},\vec{q}^{\prime}) \nonumber 
 \end{align}
The form of the Hamiltonian allows us to write it in terms of creation and annihilation operators. 

 \begin{align} \label{eqn2}
 & \hat{a} \coloneqq \frac{1}{\sqrt{2}} \left( \hat{\Pi} -i\omega\hat{\Phi} - i 
 \mathcal{L}_{\vec{q}}\hat{\Phi} \right)  ,
 &  \hat{a}^{\dagger} \coloneqq \frac{1}{\sqrt{2}} \left( \hat{\Pi} + i\omega\hat{\Phi} + i
  \mathcal{L}_{\vec{q}}\hat{\Phi}\right) 
 \end{align}
 $\omega(\phi,\vec{q} )$ is a solution of (\ref{eqn39}). 
 $\mathcal{L}_{\vec{q}}\coloneqq \vec{q}.\vec{\frac{\partial}{\partial q}}$ is directional derivative. 
$\mathcal{L}_{\vec{q}}\hat{\Phi}$ gives variation of $\Phi$ along $\vec{q}$. 
\begin{align} \label{eqn3}
  \hat{a}^{\dagger} \hat{a} = \frac{1}{2}\hat{\Pi}^{2} &+ \frac{1}{2} \left( 
  \mathcal{L}_{\vec{q}}\hat{\Phi} \right)^{2} 
 +\frac{1}{2} \omega^{2} \hat{\Phi}^{2} + i \frac{1}{2} \omega\left[ \hat{\Phi},\hat{\Pi} \right] \nonumber \\
 & -i \frac{1}{2} \left[ \hat{\Pi}, \mathcal{L}_{\vec{q}}\hat{\Phi} \right] 
 + \omega \left( \mathcal{L}_{\vec{q}}\hat{\Phi} \right)\hat{\Phi} 
\end{align}
Notice that the second term in above equation is exactly the second term in the Hamiltonian (\ref{eqn29}). Since,
\begin{equation} \label{eqn4}
 \left[ \hat{\Pi}(\vec{q}) , \mathcal{L}_{\vec{q}} \hat{\Phi} (\vec{q}^{\prime})\right] = \sqrt{\frac{h}{G}}
 \mathcal{L}_{\vec{q}} \delta(\vec{q},\vec{q}^{\prime})
\end{equation}
The forth term in (\ref{eqn3}) becomes $\mathcal{L}_{\vec{q}} \delta(\vec{0})=0$. Let us now calculate the last term 
\begin{align}
  \int d^{D}q \hspace{0.1 cm} \omega \widehat{\left( q_{i}\frac{\partial \Phi}{\partial q_{i}}\right)}\hat{\Phi}  =  
  \int d^{D}q  \frac{\partial}{\partial q_{i}}.\left( \frac{1}{2} \omega q_{i}\hat{\Phi}^{2}
  \right) 
 - \int d^{D}q \left(\vec{\frac{\partial}{\partial q}}.\left(\frac{1}{2}\vec{q}\omega \right) \right)
 \hat{\Phi}^{2} 
\end{align}
If variable $\Phi$ is chosen such that $\lim_{q \to \pm\infty} \Phi \to 0$ then total derivative term vanishes and 
(\ref{eqn3}) becomes
\begin{align} \label{eqn38}
  \hat{a}^{\dagger} \hat{a} = \frac{1}{2}\hat{\Pi}^{2} + \frac{1}{2}\left( 
  \mathcal{L}_{\vec{q}}\hat{\Phi} \right)^{2} 
 + \frac{1}{2} \left( \omega^{2}-\vec{\frac{\partial}{\partial q}}.\left( \vec{q}\omega \right) 
 \right)\hat{\Phi}^{2} \nonumber - \frac{1}{2}  \omega \delta(\vec{0}) 
\end{align}
In order to get the Hamiltonian we set 
\begin{equation} \label{eqn39}
 V= \omega^{2}-\vec{\frac{\partial}{\partial q}}.\left( \vec{q}\omega \right) 
  = \omega^{2} - D \omega - \vec{q}.\frac{\partial \omega}{\partial \vec{q}}
\end{equation}
$\omega(\phi,\vec{q})$ is a solution to the first order nonlinear differential equation.
\begin{align*}
  \hat{a}^{\dagger} \hat{a} = \frac{1}{2}\hat{\Pi}^{2} + \frac{1}{2} \left( 
  \mathcal{L}_{\vec{q}}\hat{\Phi} \right)^{2} +\frac{1}{2} V \hat{\Phi}^{2} 
  - \frac{1}{2}  \hspace*{0.1 cm} \omega(\phi,\vec{q}) \hspace*{0.1 cm} \delta(\vec{0}) 
\end{align*}
Then Hamiltonian operator can be written as 
\begin{equation} \label{eqn31}
 \hat{\textbf{H}} =  \int d^{D}q \left( \hat{a}^{\dagger} \hat{a}+ \frac{1}{2} 
 \hspace*{0.1 cm}\omega(\phi,\vec{q}) \hspace*{0.1 cm}\delta(\vec{0}) \right)
\end{equation}
Commutation relations of creation and annihilation operators can be obtained by using (\ref{eqn1}), (\ref{eqn2}) and 
(\ref{eqn4}). Only non-trivial commutation relation 
\begin{align*}
 \left[ \hat{a}(\vec{q}), \hat{a}^{\dagger}(\vec{q}^{\prime}) \right] = 
  \left(  \omega(\phi,\vec{q}) +  \mathcal{L}_{\vec{q}} \right) \delta(\vec{q},\vec{q}^{\prime})
\end{align*}
But $\mathcal{L}_{\vec{q}}\delta(\vec{q},\vec{q}^{\prime}) = \vec{q}.\vec{\frac{\partial}{\partial q}} 
 \delta(\vec{q},\vec{q}^{\prime}) = \sum_{i} q_{i}\frac{\partial}{\partial q_{i}} \delta(\vec{q},\vec{q}^{\prime})$ 
and since $x\frac{\partial}{\partial x} \delta(x)= -\delta(x)$ implying 
$\mathcal{L}_{\vec{q}}\delta(\vec{q},\vec{q}^{\prime})=-D\delta(\vec{q},\vec{q}^{\prime})$. 
\begin{align} 
 \left[ \hat{a}(\vec{q}), \hat{a}^{\dagger}(\vec{q}^{\prime}) \right] = 
  \left( \omega(\phi,\vec{q}) - D \right) \delta(\vec{q},\vec{q}^{\prime}) \label{eqn32}
\end{align} 
Rewriting Hamiltonian in terms of number operator $\hat{n} $ which gives number of field quantum. 
\begin{equation} \nonumber
 \hat{\textbf{H}} =  \int d^{D}q \hspace*{1 mm}\left( \omega(\phi,\vec{q})
 -D \right) \hat{n} + \int d^{D}q  \frac{1}{2} \hspace*{0.1 cm} \omega(\phi,\vec{q}) 
 \hspace*{0.1 cm} \delta(\vec{0}) 
\end{equation}
The Hamiltonian is self-adjoint as $\omega(\vec{q},\phi) \in  \mathbb{R} $ and a number operator $\hat{n}$ is self-adjoint
by defination. For $(\omega - D)>0$, a number operator is defined as $\hat{n}\coloneqq \hat{a}^{\dagger}\hat{a}$. Whereas for 
$(\omega - D)< 0$, it is $\hat{n}\coloneqq \hat{a}\hat{a}^{\dagger}$. Therefore the Hamiltonian is bounded from below and is 
uniquely defined as 
\begin{equation} \label{eqn33}
 \hat{\textbf{H}} =  \int d^{D}q \hspace*{1 mm} \left| \omega(\phi,\vec{q}) -D \right| \hat{n} 
  + \int d^{D}q  \frac{1}{2} \hspace*{0.1 cm} \omega(\phi,\vec{q}) 
 \hspace*{0.1 cm} \delta(\vec{0}) 
\end{equation}
\textbullet \textbf{Remarks:} \newline
The state $|n,\vec{q}>$ is an eigen state of the Hamiltonian operator. $\hat{\Phi}(\vec{q})$ is a sum of creation and 
annihilation operators, acting on the vacuum produces a field quanta with 3-metric $\vec{q}$. A role of creation and annihilation 
operator get reversed when $(\omega - D)< 0$. A field quanta (or a particle) is seen as an excitation in the combined variable 
field. The second term is vacuum energy term. \newline \newline

 \textit{Homogeneous and non-stationary space} ($q_{ab}(t,\vec{x})=q_{ab}(t)$): Since, $\partial_{a}q_{bc}=0$ 
 implying spatial curvature scalar $R^{(3)}(q_{ab},\partial_{a}q_{bc})=0$. Potential term depends only on $\sqrt{\text{det }q(t)}$ and 
 a scalar field couplings.  \newline  \newline 
 \textit{Inhomogeneous and stationary space} ($q_{ab}(t,\vec{x})=q_{ab}(\vec{x})$):
 Since 3-metric is time independent, second term in (\ref{eqn29}) vanishes. Lagrangian and Hamiltonian does not invlove integration 
 over $q$. 
\begin{equation}
 \textbf{H} = \frac{1}{2} \Pi^{2} + \frac{1}{2}V(v,R^{(3)},\phi) \hspace*{1 mm} \Phi^{2}
\end{equation}
 Riccati equation reduces to (\ref{eqn39}) reduces to $\omega^{2}=V$. 
 Hamiltonian operator for this theory is 
\begin{equation}
 \hat{\textbf{H}} = \hspace*{0.1 cm} \omega(\phi,\vec{q}) \hspace*{0.1 cm}\hat{n} + 
 \frac{1}{2}\hspace*{0.1 cm} \omega(\phi,\vec{q}) \hspace*{0.1 cm} \delta(\vec{0}) 
\end{equation}
 \section{Acknowledgment}
I am immensely grateful to Prof. Ajay Patwardhan for providing expertise. This work would not have possible 
without his guidance. Any errors are my own and should not tarnish his reputation.
 \section{Bibliography}
 $[1]$ The Dynamics of General Relativity - R. Arnowitt, S. Deser, C. W. Misner.  arXiv:gr-qc/0405109v1 19 May 2004. 
 \newline \newline
 $[2]$  Background Independent Quantum Gravity: A status report - A. Ashtekar, J. Lewandowsky.  arXiv:gr-qc/0404018v2 
3 Sept 2002. \newline \newline
 $[3]$  Gravity and the Quantum - A. Ashtekar.  arXiv:gr-qc/0410054v2 19 Oct 2004 \newline \newline
 $[4]$  Loop quantum gravity: the first twenty five years - Carlo Rovelli. arXiv:1012.4707v5 [gr-qc] 28 Jan 2012 \newline \newline
 $[5]$  Lectures on Loop Quantum Gravity - T. Thiemann arXiv:gr-qc/0210094v1 28 Oct 2002 \newline \newline
 $[6]$  Modern Canonical Quantum General Relativity - T. Thiemann Cambridge Monographs on Mathematical Physics. Cambridge 
 University Press, Cambridge, U.K., 2007. \newline \newline
\end{document}